\documentclass[nofootinbib,preprintnumbers,amsmath,amssymb]{revtex4}

\def\Re{{\rm Re}}

\usepackage{graphicx,color}
\usepackage{dcolumn}
\usepackage{bm}

\begin{document}

\title{Physically and mathematically escaping the scallop theorem}
\title{Escaping the scallop theorem}
\title{Life around the scallop theorem}
\author{Eric Lauga}
\email{elauga@ucsd.edu}
\affiliation{Department of Mechanical and Aerospace Engineering, 
University of California San Diego,
9500 Gilman Drive, La Jolla CA 92093-0411, USA.}
\date{\today}
\begin{abstract}
Locomotion on small scales is dominated by the effects of viscous forces  and, as a result, is subject to  strong physical and mathematical constraints. Following Purcell's statement of the scallop theorem which delimitates the types of swimmer designs which are not effective on small scales, we review the different ways the constraints of the theorem can be escaped for  locomotion purposes.
\end{abstract}
\maketitle

\section{Introduction}

Swimming cells, such as bacteria ({prokaryotes}) or spermatozoa ({eukaryotes}), represent the prototypical example of active soft matter. They are active as they  transform chemical energy (ATP for eukaryotes, {ion flux} for prokaryotes) into mechanical work \cite{braybook} and,  as a result, are able to continuously change shape and move in viscous environments \cite{yates86}. As mechanical entities, cells belong to the world of soft matter, displaying complex rheological properties on a range of time and spatial scales and responding to  external forcing in a time-dependent and nonlinear fashion \cite{crocker}. 

In their micron-size environment, the fluid forces  acting on swimming cells are dominated by the effect of viscous dissipation \cite{happel,kimbook}.  Seminal papers in the 1950s  laid the ground work for detailed investigations on the hydrodynamics of cell locomotion \cite{taylor51,taylor52,hancock53,gray55}, with the main goal of predicting cell kinematics, energetics, the interactions with their environment, and the general importance of fluid forces in biological form and function   \cite{lighthill75,lighthill76,brennen77,childress81,fauci06,LP09}.

In 1977, Purcell's influential paper  ``Life at low Reynolds number'' put a somewhat different spin on a field  which was already mature  \cite{purcell77}. In it, Purcell brought to light the counter-intuitive physical and mathematical constraints  arising from locomotion in an inertialess world. He demonstrated that for organisms moving in very viscous fluids, there exists a class of shape change that can never be used for locomotion, a result beautifully summarized under the name ``scallop theorem'', borrowing the name of such an  organism --- a hypothetical microscopic scallop --- which could not locomote in the absence of inertia. 

In this short review, we look back at the scallop theorem, and pose the question: 
What are the basic ingredients necessary to design swimmers able to move on small scales?
What are the different ways offered by physics to get around the constraints of the theorem?  After stating the various assumptions for the theorem to be valid (\S\ref{theorem}), we  show  how non-reciprocal shape changes (\S\ref{non-reciprocal}),  inertia (\S\ref{inertia}), hydrodynamic interactions (\S\ref{hydro}), and coupling with the physical environment (\S\ref{environment}) can all be exploited to provide locomotion on small scales.

\section{The scallop theorem}\label{theorem}

The scallop theorem has a relatively simple statement \cite{purcell77}. Consider a body changing shape in a time-periodic fashion. In the absence of inertia, the equations describing the motion of an incompressible Newtonian fluid are Stokes equations, which are linear and independent of time  \cite{happel,kimbook}. In addition, in the absence of inertia, the swimmer remains perpetually force- and torque-free \cite{LP09}. Purcell's scallop theorem can then be stated as follows.  If the sequence of shapes displayed by the swimmer is identical to the sequence of shapes displayed when seen in reverse --- so-called reciprocal motion --- then the average position of the body cannot change over one period. Another manner to describe reciprocal motion is stated  in Purcell's original  paper as:
\begin{quote}
``... I change my body into a certain shape and then I go back to the original shape by going through the sequence  in revere... So, if the animal tries to swim by a reciprocal motion, it can't go anywhere.''
\end{quote}
Time is not explicitly mentioned in the theorem and in fact, because of the linearity and  time-independence of the equations, the rate at which the sequence of shapes is being displayed is irrelevant \cite{LP09}. In Purcell's own words, 
\begin{quote}
``Fast, or slow, it exactly retraces its trajectory, and it's back where it started.''
\end{quote}
{Physically, the absence of time in the equations of motion means there is no intrinsic time scale to the swimming problem, which prevents distinguishing between forward and backward  in a reciprocal motion.}

Purcell's statements may appear simple, but are in fact far-reaching. They form the basis of a purely geometrical approach to cell locomotion  \cite{shapere87,shapere89_1,koiller96,yariv06}  and have sparked considerable attention in the area of biolocomotion from the physics and soft matter community --- so much so that  ``Life at low Reynolds number'' is now the most cited paper in the field.

The name for the theorem originates from  the simplest kind of reciprocal swimmers, namely those deforming with a single degree of freedom, such as the hinge of a hypothetical micron-scale scallop. For any swimmer with a single geometrical degree of freedom, say $\theta(t)$, then by properties of Stokes equations, its swimming speed, $u$, necessarily scales as ${u} \sim \dot \theta F(\theta)$, which is always an exact derivative, and thus averages in time to zero $\langle u \rangle = 0$. Swimmers with only one degree of freedom can thus never swim on small scales. 

Strictly speaking, the scallop theorem is valid only with the following assumptions:  a single swimmer displaying reciprocal motion in an infinite {quiescent} Newtonian fluid and in the absence of inertia {and external body forces}\footnote{{The theorem is also valid in a confinement environment as long as the boundaries display no motion.}}.  
Examining each of these assumptions in detail  suggests a way around the theorem and a design for a swimmer, which we now review. As in Purcell's original paper, we will focus only on swimming by shape change or motion and we will thus not consider chemical swimmers \cite{paxton04,golestanian05,howse07,ruckner07} or solid bodies powered by  external fields \cite{ebbens10}.

\begin{figure}[t]
\begin{center}
\includegraphics[width=0.96\textwidth]{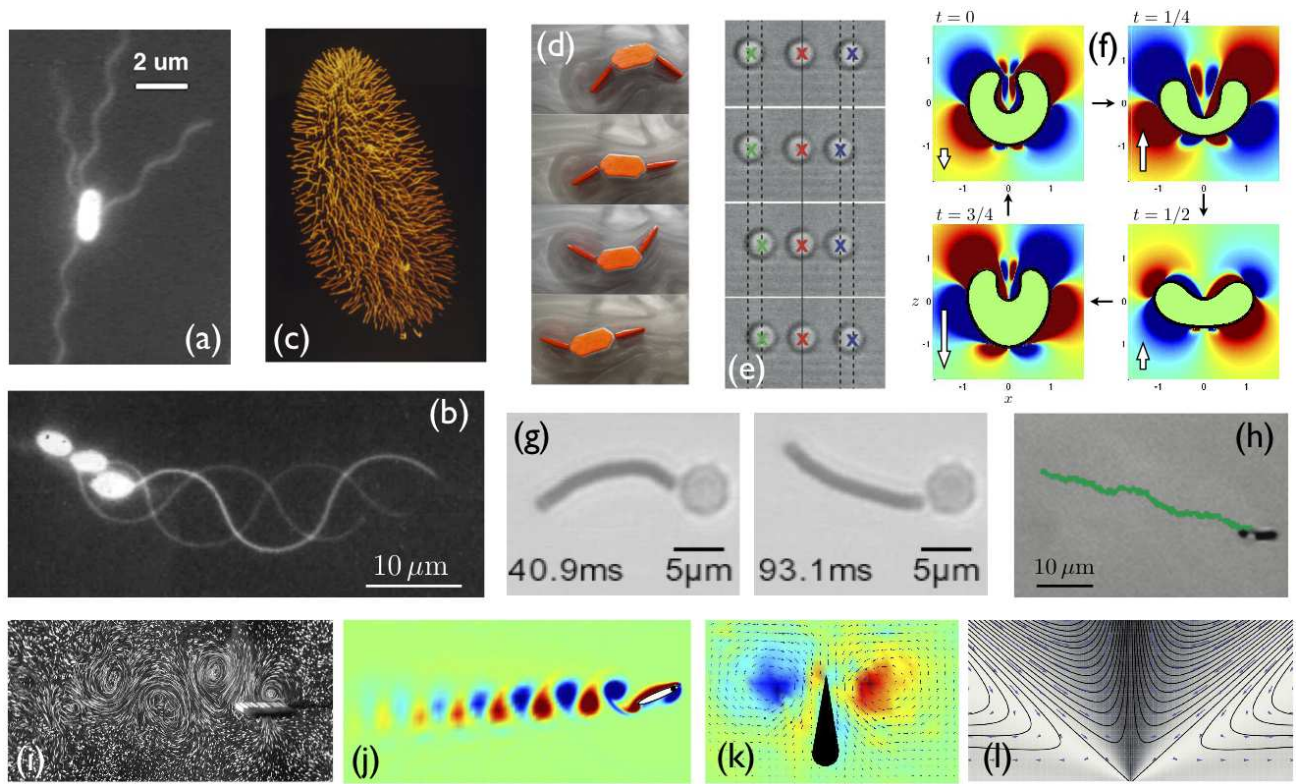}
\caption{Illustration of experimental and computational escapes from the scallop theorem. 
(a): {\it E. coli} bacterium with four helical flagella \cite{turner00}; 
(b): Superimposed pictures of a swimming spermatozoon of {\it Ciona intestinalis} \cite{brokaw65}; 
(c): {\it Paramecium} cell covered with  short cilia \cite{cnrs};
(d): Macro-scale experimental realization of Purcell's three link swimmer \cite{chanPhD}; 
(e): Micro-scale experimental realization of three-sphere swimmer using optical tweezers \cite{leoni08}; 
(f):  Computations showing the locomotion of shape-changing vesicles \cite{jellyfish}; 
(g): Microscopic realization of flexible swimming using elastic superparamagnetic filaments driven by an external magnetic field (the two images show the filament deforming at different times) \cite{dreyfus05_nature}; 
(h): Locomotion of flexible Au/Ag/Ni nanowires swimmers driven by  an external magnetic field \cite{wei10}; 
(i): Experimental demonstration that a symmetric flapping wing can undergo unidirectional locomotion if the Reynolds number is above a critical value  \cite{vandenberghe06}; 
(j): Computations showing that a flexible, flapping wing with asymmetric actuation undergoes locomotion at all finite Reynolds number \cite{spagnolie_flapping}; 
(k): Experimental measurement of the net flow and vorticity induced by a reciprocal flapper beneath a free surface \cite{trouilloud08}; 
(l): Computations for the net flow and vorticity induced by a small-amplitude reciprocal flapper in a polymeric fluid \cite{pak10}. 
All images reproduced with permission; 
(a): from  Turner, Ryu, and Berg, {\it J. Bacteriol.} {\bf 182}, 2793 (2000), Copyright 2000 American Society for Microbiology; 
(b): from C. J. Brokaw, {\it J. Exp. Biol.} {\bf 43}, 155 (1965), Copyright 1965 The Company of Biologists; 
(c): Copyright CNRS Phototh\`eque / Anne Aubusson-Fleury;
(e): from Leoni et al. {\it Soft Matt.} {\bf 5}, 472 (2009), Copyright 2009 Royal Society of Chemistry; 
(f): from Evans, Spagnolie, and  Lauga, {\it Soft Matter} {\bf 6}, 1737 (2010),  Copyright 2010 Royal Society of Chemistry; 
(g): from Dreyfus et al. {\it Nature} {\bf 437},862 (2005), Copyright 2005 Nature Publishing Group;
(h): from Gao et al.  {\it J. Am. Chem. Soc.}, {\bf  132}, 14403 (2010), Copyright 2010 American Chemical Society;
(i): from Vandenberghe,  Childress, and  Zhang, {\it Phys. Fluids} {\bf 18}, 014102 (2006), Copyright 2010 American Institute of Physics; 
 (j): from Spagnolie et al., {\it Phys. Fluids} {\bf  22}, 041903 (2010), Copyright 2010 American Institute of Physics; 
(k): from Trouilloud et al., {\it Phys. Rev. Lett.} {\bf 101}, 048102 (2008), 
Copyright 2008 American Physical Society;
 (l): from Pak, Normand, and Lauga, {\it Phys. Rev. E} {\bf 81}, 036312 (2010), 
Copyright 2010 American Physical Society.
}
\label{fig}
\end{center}
\end{figure}

\section{Non-reciprocal kinematics}\label{non-reciprocal}

\subsection{Biological swimmers: waves}

The main message of Purcell's paper is that swimmers should change their shapes in a non-reciprocal fashion. The manner in which motion occurs should thus indicate a clear direction of time, which leads naturally to the occurrence of waves. Indeed, most of swimmings cells locomote by using traveling wave-like deformation of their bodies or appendages  \cite{lighthill75,lighthill76,brennen77,childress81,fauci06,LP09}.  Swimming bacteria rotate one or more helical flagella using rotary motors embedded in the cell walls  \cite{macnab77,berg00,berg2004} leading to flagella kinematics akin to that of traveling helical  waves, and thus propulsion   \cite{chwang71,schreiner71,higdon79b} (Fig.~\ref{fig}a). Other types of bacteria swim using whole-body wave deformation propelled by flagella beneath the cell's outer membrane \cite{GoldsteinCharonKreiling1994} or  wave-like propagation of kinks in their shapes in the absence of flagella  \cite{ShaevitzLeeFletcher2005}. Spermatozoa and other singly flagellated eukaryotes swim using traveling   waves \cite{brennen77} induced  by molecular motors-driven internal sliding of polymeric filaments  inside the flagellum  \cite{summers71,brokaw89,camalet00:njp} (Fig.~\ref{fig}b). The flagella kinematics  can be planar \cite{higdon79a}, helical \cite{higdon79b}, or even doubly helical \cite{WernerSimmons2008}. The many cilia covering  some eukaryotes \cite{blake74} also deform as so-called metachronal waves  \cite{gueron97,gueron99,guirao07} (Fig.~\ref{fig}c).

\subsection{Synthetic swimmers}

Beyond the  swimming methods displayed by biological swimmers, some simpler modes of non-reciprocal motion can be devised theoretically and in the lab. 

\subsubsection{Imposing non-reciprocal kinematics}

As shown by Purcell, swimmers with a single degree of freedom cannot move. One needs therefore at least two degrees of freedom and their prescribed variation in time  should sweep a finite area in parameter space.  In his original paper, Purcell proposed such a swimmer \cite{purcell77}, namely an elongated body with two rotational hinges   \cite{becker03,tam07,avron08} (Fig.~\ref{fig}d). Subsequently, non-reciprocal swimmers of very simple shapes have been devised theoretically, including ones composed of three spheres \cite{najafi04,najafi05,dreyfus05,leoni08,golestanian10} (Fig.~\ref{fig}e), two volume-changing spheres \cite{avron05:pushme}, and two-orientation changing spheres \cite{najafi10} or ellispoids  \cite{iima09}. Beyond geometry, the two degrees of freedom could also be physical parameters, for example the volume and spontaneous curvature of a lipid vesicle  \cite{jellyfish} (Fig.~\ref{fig}f). Alternatively, the swimmer's shape and deformation change could be topologically equivalent to the inside-out rotation of a torus \cite{thaokar07,leshansky08} or  tank-treading \cite{leshansky07} for which periodicity is achieved by a continuous series of displacements  tangent to the swimmer shape. Continuous normal flows in the form of fluid jets can also be used \cite{spagnolie_jets}.

\subsubsection{Flexible swimmers: Non-reciprocal kinematics from reciprocal forcing}

A second class of simple swimmers can be designed for which a reciprocal actuation combined with flexibility or elasticity can lead to kinematics of shape change which are non-reciprocal, and thus to  locomotion.

The prototypical example of this class of swimmers is a flexible filament  actuated periodically up and down at one end where it is clamped, and free on the other \cite{WigginsGoldstein}. If the filament is rigid, its motion is reciprocal and cannot be used for propulsion. In contrast, if the filament is flexible and is actuated near the typical frequency at which viscous drag and elastic forces balance, its shape as it is actuated up (respectively down)  is concave (respectively convex), leading to non-reciprocal kinematics {and propagation of an elasto-hydrodynamic wave.}  Mathematically, the scallop theorem breaks down because time enters the problem through the viscous drag term in the equation for the filament shape (via a partial time-derivative), 
{and thus a relevant time scale can be defined.}

The generation of propulsive force and locomotion using flexibility filaments has been the center of many theoretical and computational investigations \cite{WigginsGoldstein,lowe03,lagomarsino03_filament,lauga07_pre}. A  
 macro-scale experiment confirmed the  physical picture outlined above  \cite{yu06}. Related phenomena include elastic buckling instabilities  \cite{wolgemuth00,lim04,wada06} and shape transitions  \cite{manghi06,coq08,qian08} for rotated elastic filaments. At the micro-scale, an  experimental realization of a flexible swimmer was achieved using elastic superparamagnetic filaments \cite{cebers05} actuated by external magnetic fields and attached to a red blood cell (Fig.~\ref{fig}g) \cite{dreyfus05_nature},  prompting subsequent modeling efforts \cite{roper06,gauger06,keaveny08,roper08}. {A similar implementation  was achieved using a nanometric silver filament attached to an externally-driven  ferromagnetic nickel head  (Fig.~\ref{fig}h) \cite{wei10}. 
In all these cases however, it is the presence of  external torques (via external magnetic fields) that allows locomotion, and thus they do  not represent  true self-propelled motion.}


\section{Inertia}\label{inertia}

For the scallop theorem to be valid, all inertial terms in the equation of motion of the swimmer should be set to zero.  Naturally, they cannot exactly disappear unless no motion occurs, and thus a fundamental question arises, namely how much inertia is needed to escape the constraints of the theorem? Is the scallop theorem valid only asymptotically, or does it stand as long as inertia is below a certain limit? 
These questions were first posed by Dudley and Childress \cite{childress04} who studied the behavior of a mollusk able to use both reciprocal and non-reciprocal modes of locomotion, and who  postulated that a finite amount of inertia was necessary for locomotion  to be able to occur. 

Mathematically, three qualitatively different Reynolds numbers can be defined. Consider a swimmer of typical size $L$ and density $\rho_s$ undergoing reciprocal motion of amplitude $A$ and frequency $\omega$ in a Newtonian fluid of density $\rho$ and shear viscosity $\mu$. Using a typical velocity scale $U\sim A \omega$, the natural Reynolds number for the reciprocal motion is given by  $\Re=\rho L A \omega/\mu$, and  is the one corresponding to the nonlinear advection term in the Navier Stokes equations {(for example, in water,  $\Re \approx 10^{-4}$ for {\it E. coli} while $\Re \approx 10^{-2}$ for human spermatozoa)}. The oscillatory Reynolds number, corresponding to the linear unsteady Stokes term, is given by $\Re_\omega=\rho L^2 \omega/\mu$. Finally, the Reynolds number based on the body inertia is $\Re_s=\rho_s L^2 \omega/\mu$,  sometimes called a Stokes number, which quantifies the typical ratio between the rate of change of the swimmer momentum and  the magnitude of the viscous forces in the fluid.

For small amount of inertia, the breakdown of the scallop theorem  occurs either continuously or discontinuously with these Reynolds numbers  depending on the geometrical symmetries in the reciprocal actuation. In the case of symmetric shapes --- typically simple flappers --- experiments and modeling demonstrated that a finite, order one, amount of inertia is necessary, indicating a discontinuous transition through an inertial hydrodynamic instability  \cite{childress04,vandenberghe04,alben05,lu06,vandenberghe06,childress_conf} (Fig.~\ref{fig}i). As a difference, in the case of asymmetric shapes or actuation, the transition is continuous, with locomotion occurring either as some power of $\Re$ \cite{lauga_purcell} or both $\Re$ and $\Re_\omega$ (with $\Re/\Re_\omega$ constant) \cite{spagnolie_flapping,roperPhD} (Fig.~\ref{fig}j). Interestingly, for asymmetric shapes, a continuous transition with swimmer inertia was  obtained in the absence of  fluid inertia  ($\Re=\Re_\omega=0$), with locomotion occurring as powers of $ \Re_s$
\cite{Gonzalez-Rodriguez}.

\section{Hydrodynamic interactions}\label{hydro}
 
The inertialess scallop envisioned by Purcell as the prototypical non-swimmer is isolated in the fluid.  It turns out however that hydrodynamic interactions with other such non-swimmers, {or more generally flexible entities,} can be exploited to swim. Physically, as cells or other synthetic swimming devices do work on the surrounding fluid,  they act as hydrodynamic disturbances on the otherwise-quiescent environment, thereby  setting up flow fields which are in general dipolar  \cite{LP09}. In biology these flow fields have important consequence on the generation of  collective modes of locomotion \cite{mendelson99,wu00,dombrowski04,cisneros07,sokolov07} and rheology at the whole-population level \cite{sokolov10,rafai10}.

Although a body undergoing reciprocal motion cannot swim, two bodies undergoing reciprocal motion with nontrivial phase differences  are able to take advantage of the unsteady hydrodynamic flows they create to undergo nonzero collective and relative dynamics; there is thus no many-scallop theorem  \cite{laugabartolo08,alexander08}. As each reciprocal swimmer behaves in general as an unsteady dipole, the collective effect arises from the time-rectification of such unsteadiness, and thus decays generically as $1/d^3$, where $d$ is the typical swimmer-swimmer distance (or even faster if additional geometrical symmetries are present \cite{laugabartolo08,BL10}). Naturally, two reciprocal non-swimmers taken as a whole are not unlike a single non-reciprocal swimmer, although the qualitative details of their locomotion do differ \cite{laugabartolo08}.

{Experimentally, this effect was demonstrated for hydrodynamic interactions between a rigid flapper, beating in a reciprocal fashion, and a flexible boundary (free surface). The rectification of the reciprocal  flow by the free surface motion leads to flow and forces scaling quadratically with the applied flapping frequency, and the creation of a reciprocal pump   \cite{trouilloud08} (Fig.~\ref{fig}k).
The experimental application of these ideas to a collection of free-swimming bodies  remains however to be confirmed}. To generate reciprocal motion with nontrivial phase-differences, one possibility would be to use elastic field-responsive particles under a uniform AC forcing;  particles with different relaxation times would respond to fields with different phases, and thus would be able to move collectively \cite{laugabartolo08}. In the case of purely identical non-swimmers,  two of them cannot swim, but three or more are able to move   \cite{BL10}. In that case, the phase  differences in body kinematics are induced by hydrodynamic flows, leading to a slow $1/d^7$ effect  \cite{BL10}.

\section{Physical environment}\label{environment}

In the scallop theorem, the assumption that locomotion takes place in  a Newtonian environment  is crucial, as it allows the inertialess equations of fluid motion to be linear and independent of time. A change of the mechanical and rheological properties  of the fluid would however naturally lead to a different type of conclusion. Complex fluids are abundant in biology,  and cell locomotion often takes place in strongly elastic polymeric fluids  \cite{shukla78,katz80,suarez92}, which has been the focus of much recent work  \cite{lauga07,FuPowersWolgemuth2007,FuWolgemuthPowers2008,FuWolgemuthPowers2009,laugaDe,teran_PRL}.

As the fluid becomes non-Newtonian, three different physical effects can potentially be exploited to generate small-scale locomotion \cite{birdvol1,birdvol2}. First, complex fluids possess in general rheological properties which are rate dependent.  {In particular, viscosities often display shear-thinning behavior, meaning they decrease with shear rates}. In this type of fluid, and in contrast with the Newtonian case, the rate at which the reciprocal sequence of shapes  is being displayed would matter, a result which could be used to design a reciprocal swimmer. This was recently demonstrated theoretically for bodies swimming using a reciprocal  helical actuation at different rates in model polymeric fluids \cite{FuWolgemuthPowers2009}.

The second physical effect to be exploited is that of normal stress differences, which arise from the stretching by the flow of the  microstructure suspended in the complex fluids. Normal stress differences scale quadratically with the applied shear \cite{birdvol1} and remain thus identical under a reversal a time, allowing propulsion. Locomotion using normal stress differences was demonstrated theoretically for a three-dimensional body undergoing small-amplitude reciprocal motion  at constant rate \cite{laugaDe}. The generation of forces  and flow by reciprocal flapping was also reported   \cite{normand08,pak10} {(Fig.~\ref{fig}l). }

The last physical effect to be exploited is that of stress relaxation.   Even for small-amplitude motion and linearized dynamics, the simplest evolution equation for the stress in a polymeric fluid contains a memory term in the form of a partial time derivative times a relaxation time. Whether, even in the linear regime, stress relaxation can be taken advantage of for locomotion purposes is an intriguing,  but yet unexplored, possibility.

\section{Conclusion}

In this short review, we have used Purcell's scallop theorem as a framework to lay out the basic physical principles behind the design of small-scale swimming devices. We have shown how non-reciprocal kinematics, inertia, hydrodynamic interactions, and the nature of surrounding environment  can all be physically exploited to achieve small-scale propulsion.  {With advances in micro- and nano-fabrication, the discussion on the theorem can now move from that akin to a mathematical exercise to a true engineering challenge.}

As briefly mentioned in Ref.~\cite{childress81}, there exists at least another class of body motion which always leads to zero locomotion in a Newtonian fluid, namely those for which the time-reversal of the motion  is identical to its mirror-image (for example, the motion of a rod sweeping the envelope of a cone). The formal  derivation of the complete class of non-swimming body kinematics would provide a new thrust in small-scale locomotion research by allowing novel opportunities to get around these mathematical constraints.


\section*{Acknowledgements}
Discussions with Denis Bartolo are gratefully acknowledged.  I thank B. Chan, A. E. Hosoi, A. Aubusson-Fleury and the CNRS Phototh\`eque for providing me with the images reproduced in Fig.~\ref{fig}c and d, 
as well as  the authors from Refs.~\cite{turner00,brokaw65,chanPhD,leoni08,jellyfish,dreyfus05_nature,wei10,vandenberghe06,spagnolie_flapping,trouilloud08,pak10} who gave us permission to reproduce their images. This work was supported in part by the US National Science Foundation through grant number CBET-0746285. 

\bibliography{bib_2010}
\end{document}